\documentclass[letterpaper]{article} 
\usepackage{aaai23} 
\usepackage{times}  
\usepackage{helvet}  
\usepackage{courier}  
\usepackage[hyphens]{url}  
\usepackage{graphicx} 
\usepackage{natbib}  
\usepackage{caption} 
\usepackage{algorithm}
\usepackage{algorithmic}

\usepackage{amsmath}
\usepackage{refcount}
\usepackage{longtable}
\usepackage{subcaption}
\usepackage{multirow}
\usepackage{tablefootnote}
\usepackage{booktabs}
\usepackage{xcolor}
\usepackage{makecell}
\usepackage[title]{appendix}

\usepackage{newfloat}
\usepackage{listings}
\DeclareCaptionStyle{ruled}{labelfont=normalfont,labelsep=colon,strut=off} 
\floatstyle{ruled}
\newfloat{listing}{tb}{lst}{}
\floatname{listing}{Listing}
\title{Topic Shifts as a Proxy for Assessing Politicization in Social Media}
\author{
    Marcelo Sartori Locatelli,
    Pedro Calais,
    Matheus Prado Miranda\equalcontrib, João Pedro Junho\equalcontrib, Tomas Lacerda Muniz, Wagner Meira Jr., Virgilio Almeida\\
}
\affiliations{
    Universidade Federal de Minas Gerais, Belo Horizonte, Brazil\\

    \{locatellimarcelo, pcalais, matheus.prado, joaopedro.junho, tomas.muniz, meira, virgilio\}@dcc.ufmg.br
}

\usepackage{bibentry}

\begin{document}

\maketitle

\begin{abstract}

Politicization is a social phenomenon studied by political science characterized by the extent to which ideas and facts are given a political tone. A range of topics, such as climate change, religion and vaccines has been subject to increasing politicization in the media and social media platforms. In this work, we propose a computational method for assessing politicization in online conversations based on \emph{topic shifts}, i.e., the degree to which people switch topics in online conversations. The intuition is that topic shifts from a non-political topic to politics are a \emph{direct} measure of politicization -- making something political,  and that the more people switch conversations to politics, the more they perceive politics as playing a vital role in their daily lives. A fundamental challenge that must be addressed when one studies politicization in social media is that, a priori, \emph{any} topic may be politicized. Hence, any keyword-based method or even machine learning approaches that rely on topic labels to classify topics are expensive to run and potentially ineffective. Instead, we learn from a seed of political keywords and use Positive-Unlabeled (PU) Learning to detect political comments in reaction to non-political news articles posted on Twitter, YouTube, and TikTok during the 2022 Brazilian presidential elections. Our findings indicate that all platforms show evidence of politicization as discussion around topics adjacent to politics such as economy, crime and drugs tend to shift to politics. Even the least politicized topics had the rate in which their topics shift to politics increased in the lead up to the elections and after other political events in Brazil -- an evidence of politicization. The code is available at \url{https://github.com/marceloslo/Topic-Shifts-as-a-Proxy-for-Assessing-Politicization-in-Social-Media}.
\end{abstract}

\section{Introduction}

\label{sec:intro}
Nowadays, any person may publicly share their views on a given subject with a far larger reach than they would have otherwise~\cite{boynton2016agenda}, 
social media platforms have enabled a plethora of studies in the social sciences and, more specifically, in the political sciences~\cite{ computational_social_science, edelman_computational_2020}.
While threats to the validity of studies based on social media data are still a concern~\cite{validity}, access to large amounts of digital behavioral data has allowed political scientists to pair with their computer science peers to study the role of social media in government behavior~\cite{social_media_communications}, voter engagement~\cite{GROVER2019438}, news coverage and its bias~\cite{new_media_bias, political_news_coverage} and even election forecasts~\cite{election_forecasts}. 
More specifically, two widely recognized political processes received special attention with respect to how they shape (and are shaped) by social media, namely, \emph{polarization} and \emph{politicization}.


While polarization refers to the process by which two or more political groups selectively chooses to consume opinions they already agree with and adopt increasingly extreme and antagonistic viewpoints~\cite{layton2021demographic}, politicization, the focus of our work, is the act of marking or naming something as political~\cite{wiesner2021rethinking}. Topics recently subject to increasing politicization include  climate change~\cite{politicization_climate_change}, COVID-19~\cite{Politicization_and_Polarization_COVID}, religion~\cite{politicization_religion},  and culture and science in general~\cite{politicization_culture,  politicization_science}. By adding an ideological charge to a non-political issue, politicization  may lead to manipulation, increased hostility and a lack of trust to the public debate. Next we present a concrete example of politicization in a news article published by Folha de São Paulo newspaper in Twitter during the Brazilian 2022 presidential elections:

\begin{quote}{\textbf{Post}: ``Latin America: Evangelical gays defy churches and get married after pro-LGBTQIA+ referendum in Cuba."\footnote{\label{translate}Posts have been translated from Portuguese to English. The comment was paraphrased to protect the identity of the user.}}\end{quote}
\begin{quote}{\textbf{Comment}: ``Recently, @folha's headlines have seemed to be made specifically to be used by Bolsonaro's supporters and fuel disinformation."\textsuperscript{\getrefnumber{translate}}}\end{quote}

Note that the original post touches on an (apolitical) religious and LGBTQIA+ topic that was quickly labeled as politically motivated -- in particular, meant to be politically exploited by supporters of Brazilian 2022 presidential candidate Jair Bolsonaro. In this work, we devise a computational method that directly models two key aspects that characterize politicization:

\begin{enumerate}
    \item The transition from a non-political to a political topic.
    On unfiltered social media datasets comprising conversations on several topics, the post starting the discussion and the
    comments written in reaction to it can be both political and non-political; to detect politicization, we find \emph{topic shifts}~\cite{sun2019topic} in which the
    original post is non-political, but comments are political, as a proxy for politicization.
    \item The fact that, a priori, \emph{any} non-political topic may be politicized, and, hence, we cannot predict or anticipate all topics that may become political; manual labeling
    of individual posts that cover the wide spectrum of non-political topics would be costly. We  have a set of high-precision positive labeled posts and comments derived from unambiguous political keywords, but we do not have negative (non-political) labels.  
    To address this challenge, we resort to a semi-supervised machine learning strategy that learns from positive and unlabeled examples known as Positive-Unlabeled Learning (PU learning for short)~\cite{learning_from_positive_unlabeled}. The key capability of PU learning is that
    it works in the absence of
    negative training examples  and finds a boundary between positive and (hidden) negative examples under the assumption that their feature distribution is different.
\end{enumerate}

Our method extends existing strategies to study politicization in online media, which typically share two limitations: they are focused on a single topic and are fully keyword-based or require negative (non-political) labels, which limits the extent to which topic shifts can be observed. By starting with a small seed of high-precision political keywords, but expanding them through a two-step PU Learning strategy, we were able to perform a general and broad characterization of politicization on social media which found, based on Twitter, YouTube, and TikTok data collected during the 2022 Brazilian presidential elections, that:

\begin{itemize}
\item By starting with a small seed of high-precision political keywords and using word2vec features in an XGBoost classifier, it is possible to reach a 86\% F1 score to distinguish between political and non-political news posts and comments;

\item Politicization is a widespread phenomenon on social media. While political content is more common in Twitter and less prevalent on TikTok, in all three platforms, at least one out of two non-political news posts will receive at least one political comment. 

\item Topics that are more heavily politicized include hard news such as the economy, media behavior, education, and drugs, but even soft news such as sports are politicized to some extent. Even for the least political topics, the rate of which their topic was shifted increased considerably during and after the Brazilian elections, suggesting politicization.

\end{itemize}


Our paper is organized as follows. 
Initially, we discuss related work on politicization and provide more details on how our research extends the existing literature.
Next, we detail the datasets we use and the computational method we employ to find news articles that are highly politicized. Then, we use the model to characterize politicization along several dimensions, such as prevalence, topics, and time. Finally, we discuss conclusions and future research directions.

\section{Related Work}
\label{sec:related_work}
\begin{table*}[ht]
\small
\centering
\begin{tabular}{lrrrrc}\toprule
                 & \textbf{\# News Sources} & \textbf{\# Posts} & \textbf{\# Avg. Comments per Post} & \textbf{\# Avg. Comments per User} &\textbf{Data Collection Period}\\\midrule
\textbf{TikTok}  & 41 & 8,814 & 20.28 & 1.37  & 2022-08-24 to 2022-11-01   \\
\textbf{Twitter} & 50 & 119,691 & 27.75 & 5.66  & 2022-08-26 to 2023-03-03 \\

\textbf{YouTube} & 43 & 12,616 & 347.80 & 4.61   &2022-01-01 to 2023-05-06 \\\bottomrule
\end{tabular}

\caption{Statistics per social media platform. Based on profiles of relevant news sources in Brazil, we collected their posts and the comments posted by other users in reaction to each post.}
\label{tab:statistics}

\end{table*}


Politicization affects how individuals allow their motivations and emotions to influence how they interpret new information~\cite{taber2009motivated}, and political science has deeply studied politicization through online social media data. Observational studies usually focus on a single non-political topic, which can be as specific as the adoption of a low-carb diet in Sweden~\cite{1ow_carb_diet}, a Star Wars movie~\cite{bay2018weaponizing}, or a mega sports event such as the World Cup~\cite{politicization_world_cup}; 
typical conclusions are that those topics have been subject to increasing politicization.

One can see if a non-political topic is made political by correlating it with polarization~\cite{weber2013inferring,fake_news_policitization, Stick_To_Sports}:
if distinct political groups refer to a non-political topic differently or at different rates, it is a strong signal of a politicized topic, such as ``gun violence" and ``religious freedom" receiving different attention from Democrats and Republicans~\cite{communities_shape_political_beliefs}. Another common strategy to evaluate politicization is to count the extent to which a piece of content mentions political actors~\cite{politicization_polarization, Politicization_and_Polarization_COVID};  the more a non-political content (such as COVID-19) is linked to politician names or political concepts, the more politicized it  is.

We generalize existing politicization studies over social media data in two important directions. First, instead of focusing on a specific topic such as COVID~\cite{Politicization_Ivermectin}, we enable the study of politicization in general social media data that comprises \emph{both} political and non-political news articles and associated comments. Hence, we are able to assess the general prevalence of politicization in online spaces. To that end, we employ a classification approach fed with high-precision political keywords. To avoid the need of labeling non-political content, we employ a semi-supervised approach called Positive-Unlabeled (PU) Learning.
PU Learning has been used to learn from social media posts in the context of tasks such as  sentiment analysis~\cite{wang2017sentiment}, classification of user profiles~\cite{teacher_pu_learning}, and fake news detection~\cite{fake_news_early_detection}. Here, we use the technique to learn to classify unlabeled posts and comments, which can be either political or non-political. 

Current research also focuses on inferring politicization by examining whether a piece of non-political content is \emph{intermixed} with political content. For instance, the mention to political actors in a news article about COVID or political content being posted in a non-political community~\cite{politicization_polarization, wojcieszak2009online}. In contrast, we investigate a more explicit phenomenon: the actual \emph{transition} of previously non-political content into a political context, as exemplified by the shift of the LGBTQIA+ topic to politics in the Introduction. Our method perceives politicization as more of a \emph{process}, given that in a topic shift, the original post was either less political or entirely apolitical. Our modeling approach is more aligned with the perspective that politicization involves the transformation of previously non-political matters into political ones~\cite{politicization_compared}, since we look for a non-political content which is followed, in a distinct action initiated by another person in the future, by a political comment.

\section{Dataset: YouTube, Twitter and TikTok}
\label{sec:data}

Motivated by the highly polarized 2018 Brazilian presidential elections~\cite{polarization_Brazilian_Elections,layton2021demographic} and the extensive use of social media platforms by the presidential candidates, we study the politicization of news posts in the context of the 2022 elections. We collected data from three platforms: YouTube, Twitter, and TikTok. While the first two have been the target of a plethora of studies over the years~\cite{montag2021psychology,ling2022slapping}, the latter is a platform that has grown extremely fast recently, and, as such, 
the behavior of its users is still poorly understood by the research community \cite{medina2020dancing}.

We collected all posts (and associated comments) published by popular Brazilian news sources
(or their equivalents on a given platform) and the reactions to them (likes, shares, replies, and comments).  Therefore, we observe not only political but also non-political news and associated comments, which enables a range of new perspectives on political behavior, including observing politicization, a concept that by its nature requires non-political data to be appropriately observed. To allow for an appropriate analysis, only comments with more than 5 tokens were considered.


The selection of news source profiles was conducted in order to select some of Brazil's most prominent digital news sources.
Therefore, the profiles with the most significant engagement, measured in followers or likes, were those picked for the research. Note that not all of them have profiles on each of the three platforms.

On YouTube and Twitter, these were collected using the official APIs, YouTube Data API v3 and Twitter API v2, respectively, made available by the platforms. On TikTok, due to the lack of an official API at the time of the collection period, an unofficial API\footnote{https://github.com/davidteather/tiktok-api} as well as web scrapers were used. Table \ref{tab:statistics} shows the statistics for the collected data. We aimed to collect news on all platforms well into 2023. However, due to TikTok updates in November 2022, the unofficial API and other ways of collecting data stopped working, interrupting the collection on that platform.

\section{Detecting Politicization With Positive-Unlabeled Learning}
\label{sec:pu_learing}
Manually inspecting or labeling posts in the datasets searching for political content and politicization of non-political content would be costly and time consuming. Differently from~\cite{rajadesingan2021political}, which focus on Reddit communities, we do not have social groups we could label as political or non-political, since, on YouTube, Twitter,  and TikTok, the discussion is centered around individual content and not communities. However, in the context of (Brazilian) politics, it is fairly easy to identify some
posts that are very unambiguously political, such as a post that
cites Lula or Bolsonaro (the front-runners of the 2022 presidential elections). Additionally, in our dataset, especially on
TikTok, \emph{\#eleicoes2022} was one of the most prevalent hashtags, referring to the presidential elections.
Given these assumptions, by using the 2 most prominent presidential candidates (Lula and Bolsonaro) as well as the aforementioned hashtag, it was possible to identify a positive set P composed of news and comments very strongly linked to politics.

By using those high-precision political keywords, we can conduct a first examination of the prevalence of political news posts and associated comments. In Tables~\ref{tab:politicalratio1} and~\ref{tab:politicalratio2}, we observe that at least 26\%, 19\% and 14\% of news posts on YouTube, Twitter and TikTok are unambiguously political, respectively. In political news posts, we see that comments that contained the aforementioned political keywords were between 2 to 6 times more frequent, when compared to the unlabeled posts, which is consistent with the expectation that political news attract more political comments. Due to the fact that these numbers were obtained from a small number of keywords, the amount of political content is actually higher than that, as we will discuss in the Characterizing Politicization section.
The interesting numbers, however, are those in Table~\ref{tab:politicalratio2}: among unlabeled news posts, between 5 and 8\% of the comments are political, and from 22 to 49\% of comment threads contain at least one post that is unambiguously political, which may hide the politicization of non-political news posts.

\begin{table}[tbhp]
\small
\centering
\begin{tabular}{cccccc}
\toprule
\multicolumn{1}{c}{} & \multicolumn{1}{c}{} & \multicolumn{3}{c}{\textbf{Comments in Political News Posts}} \\
\cmidrule(rl){3-5}
\textbf{Platform} & \textbf{P Posts} & {P} & {U} & {At least one P Comment} \\
\midrule
YouTube & 26\% & 27\% & 73\% & 94\% \\
Twitter & 19\% & 15\% & 85\% & 78\% \\
TikTok & 14\% & 31\% & 69\% & 85\% \\
\bottomrule
\end{tabular}
\caption{Ratio of political (P) posts per platform and prevalence of political comments. Since we considered high-precision political keywords, these are  approximate lower bounds for the prevalence of politics in the dataset.}
\label{tab:politicalratio1}
\end{table}

\begin{table}[tbhp]
\small
\centering
\begin{tabular}{cccccc}
\toprule
\multicolumn{1}{c}{} & \multicolumn{1}{c}{} & \multicolumn{3}{c}{\textbf{Comments in Unlabeled News Posts}} \\
\cmidrule(rl){3-5}
\textbf{Platform} & \textbf{U Posts} & {P} & {U} & {At least one P Comment} \\
\midrule
YouTube & 74\% & 8\% & 92\% & 49\% \\
Twitter & 81\% & 6\% & 94\% & 32\% \\
TikTok & 86\% & 5\% & 95\% & 22\% \\
\bottomrule
\end{tabular}
\caption{Ratio of unlabeled (U) posts per platform and prevalence of political comments. Unlabeled posts can be either political or non-political.}
\label{tab:politicalratio2}
\end{table}



\subsection{Two-Step PU Learning} 

To actually assess politicization, we must be able to classify unlabeled content as political or non-political to some degree. Given a set P of positive examples about politics
and a set U of unlabeled examples (which contain hidden examples about politics and content that is non-political), we want to build a classifier using P and U that can identify positive (political) and negative (non-political) documents in U~\cite{liu_web_data_mining}. 

To operate in this semi-supervised setting, we employ a Positive-Unlabeled (PU) Learning strategy called \emph{two-step}~\cite{learning_from_positive_unlabeled}. In this strategy, we first (1) find reliable negative examples (non-political examples) and then (2) use supervised or semi-supervised techniques with the labeled, reliable negatives and, optionally, unlabeled examples as inputs. The underlying assumption is that unlabeled positive examples are similar to their labeled counterparts, while negative examples are sampled from a different distribution.

\begin{figure*}[ht]
    \centering
   \includegraphics[width=0.75\textwidth]{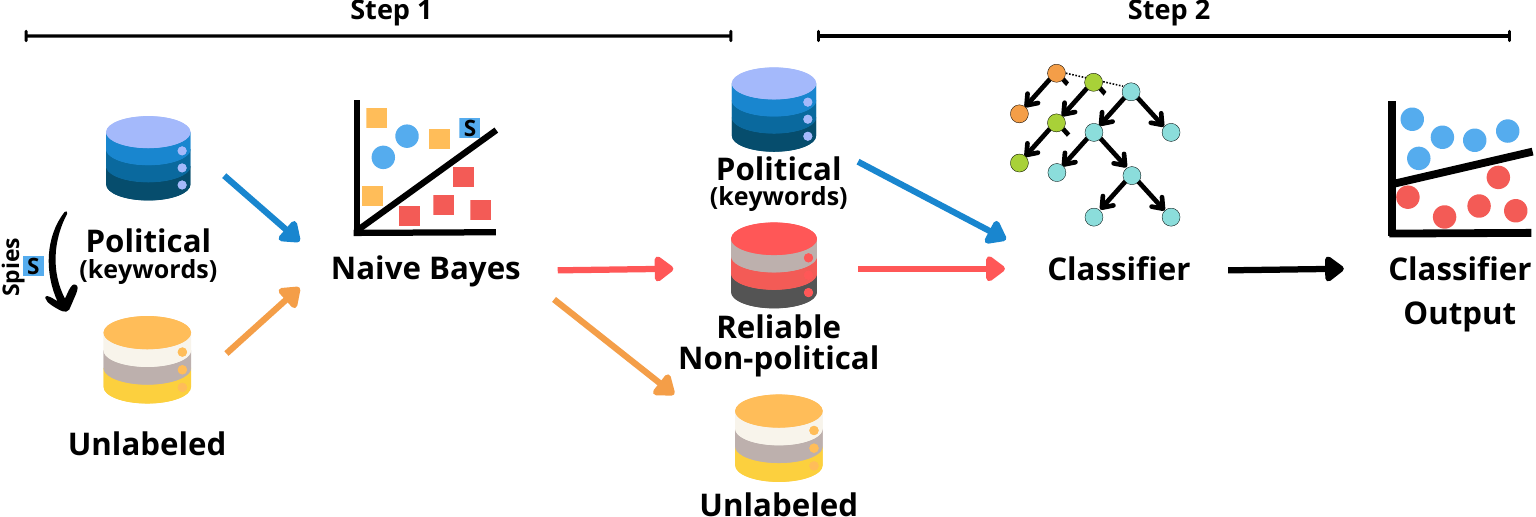}
   \caption{The two-step Positive-Unlabeled (PU) learning technique.
   Step 1 is fed with political and unlabeled examples and divides the unlabeled set into two sets -- reliable non-political and a smaller unlabeled set. Step 2 is a traditional binary classifier fed with political and reliable non-political examples.  
   Squares represent examples treated as unlabeled during the first step, while circles represent examples treated as labeled. Red represents examples classified as non-political, blue represents political and yellow, unlabeled.}
  \label{fig:flowchart}
\end{figure*}

\textbf{First PU Learning step: extracting reliable negative examples with spies.}
To find reliable non-political (negative) examples, we use \emph{spies}~\cite{liu2002partially}.
Spies are a random selection of a fraction
of the positive labeled examples (we used s\% = 10\% of positive examples), which will be treated as unlabeled examples. Since we know they are actually positive examples, we will use the probability scores attributed to the spies by the classifier trained on P (with spies removed) and U to calibrate the label probability that delimits the boundary that separates positive from negative examples -- see Step 1 depicted in Figure~\ref{fig:flowchart}.

In an ideal world, we would classify the reliable negatives as the examples that were attributed probabilities lower than $\min{P[c=P|s_1],P[c=P|s_2]...P[c=P|s_k]}$, where $s_k$ is the k-th spy and $c$ is the predicted class, we call this threshold $t$. However, due to the existence of noise and outliers, some spies may have lower probability than most negative documents, so a noise level $l$ is used to estimate $t$ so that $l\%$ of the documents have probability lower than $t$. We used $l=15\%$ following advice from~\cite{liu2002partially} that any rate between 5 and 20\% works well. 

The pseudo-code for this step is shown in Algorithm~\ref{alg:spies}. We used TF-IDF as features and a Naive Bayes Classifier. 

\begin{algorithm}
\caption{PU Learning Step 1 algorithm}\label{alg:spies}
\begin{algorithmic}
    \STATE $N \gets \emptyset$  \COMMENT{Initialize Reliable Negatives}
    \STATE $S \gets$ sample($P,s\%$)    \COMMENT{Initialize Spies}
    \STATE $US \gets U \bigcup S$
    \STATE $P \gets P-S$
    \STATE Train Naive Bayes using $US$ and $P$
    \STATE Classify each document in $US$
    \STATE Estimate $t$ using S
    \FOR{$u_i$ in $U$}
        \IF{$P[c=P|u_i] < t$}
            \STATE $N \gets N \bigcup \{u_i\}$
            \STATE $U \gets U - \{u_i\}$
        \ENDIF
    \ENDFOR 
\end{algorithmic}
\end{algorithm}

\textbf{Second PU Learning step: Traditional binary classifier.} In the second step,
we learn a traditional classifier fed with positive and the negative examples obtained from step 1 -- see Figure \ref{fig:flowchart}.
 We tested a variety of word representations and classifiers, including a fine-tuned BERT model, however, our discussions will focus on word2vec as the word representation and gradient boosting as the classifier, using the XGBoost library~\cite{Chen:2016:XST:2939672.2939785}. 

 \textbf{Baselines}. To appropriately factor in the impact of the 2-step PU Learning solution for our political classification problem, we compare it with a few baselines:

 \begin{enumerate}

 \item A keyword-based classifier based on the political keywords to expose the extent to which a simple match of keywords is enough to separate political from non-political content;
 \item A gradient boosted tree classifier that considers all unlabeled content as negative, 
 to assess the actual need for treating unlabeled examples in a PU fashion;
 \item A PU Learning strategy based on the incorporation of class priors to calibrate the classification\footnote{Based on this implementation: \url{github.com/pulearn/pulearn.}}~\cite{elkan2008learning}.
  This strategy, in principle, should not be adequate for our problem since the high-precision keywords used to create P tend not to be a random sample of the full set of positive examples but rather a biased and easier-to-classify sample.
\end{enumerate}

\section{Experiments}
\label{sec:experiments}
To assess the capacity of PU Learning in distinguishing political from non-political content, we manually annotated 1,500 news posts and an equal number of comments\footnote{Among these, 500 were posted on Twitter, 500 on YouTube, and 500 on TikTok.}. Three of the authors independently assigned labels, and we conducted a majority vote. The annotators identified explicit or implicit mentions of political candidates, parties, affiliated public offices, and notable political groups and supporters. Discussions related to specific legislative actions, state policies, calls for political engagement, as well as political protests were also categorized as political content. For the purpose of this study, scenarios involving foreign politics were considered as \emph{non-political} due to our focus on the Brazilian elections context.  62\% of the news posts and 55\% of the comments were labeled as political, respectively.


We utilized Fleiss' Kappa~\cite{KochLandis77} as a measure of inter-rater reliability for the resulting labels, and we obtained a score of 0.748. This score falls within the ``substantial agreement" range, indicating good consistency in the labels provided by the annotators.

\begin{figure}[bt]
    \centering
    \includegraphics[width=0.9\columnwidth]{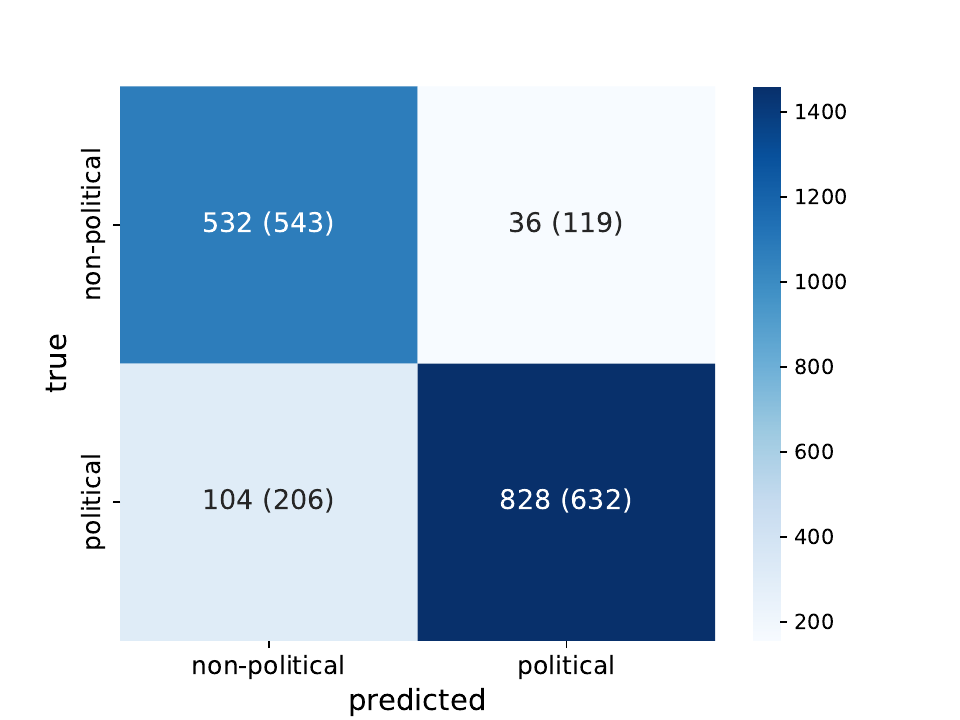}
    \caption{Confusion matrix for news and comments predictions for the XGBoost PU learning model. The numbers inside the parentheses show the comment predictions, while those outside show news post predictions. Performance is superior for news posts, possibly due to comments having less context and structure than well-formed news headlines. Note how the errors are relatively well balanced.}
    \label{fig:enter-label}
\end{figure}

We compare the performance of the baseline models and the PU variants: the political keyword classifier, an XGBoost that naively treats unlabeled examples as negative, and three flavors of PU Learning: one based on using class priors and two using the 2-step strategy, one using a fine-tuned PT-BR BERT model~\cite{souza2020bertimbau}, and the other using XGBoost, however the latter heavily outperformed the former. 
Hyper-parameters were tuned using random search. Accuracy, weighted average, recall, precision, and F1 for all models are summarized in Table \ref{tab:results}, and we show the confusion matrix for the best model(two-step PU XGBoost) for news posts and comments in Figure \ref{fig:enter-label}.

The 2-step PU method with XGBoost is the best performing model. F1 score is 0.86, which is aligned with similar works that involved the use of two-step PU learning techniques in other contexts involving text~\cite{li2014spotting,fusilier2015detecting}. Interestingly, all models perform
better for the news posts (up to 0.92 F1) when compared with comments (up to 0.80 F1), which is expected as their text is not only longer, on average, than comments, but also has more context and structure.


    


\begin{table}[bt]
\small
        \centering
\begin{tabular}{lrrrrr} \toprule
\textbf{Model}                 & \textbf{ACC} & \textbf{F1} & \textbf{Recall} & \textbf{Precision} \\ \midrule
\textbf{Keyword-based}          & 0.65              & 0.55        & 0.40            & \textbf{1.00}               \\
\textbf{XGBoost w/ unlabeled} & 0.73              & 0.71        & 0.58            & 0.94               \\
\textbf{Class Prior XGBoost}    & 0.80              & 0.82        & 0.79            & 0.86               \\
\textbf{Two-step XGBoost}       & \textbf{0.84}              & \textbf{0.86}        & \textbf{0.82}           & 0.90                        \\\bottomrule   
\end{tabular}    

    \caption{Average scores for each model considering news and comment predictions. Values in bold represent the best performing model for a given metric. In comparison to a simple keyword-based approach, models expand recall while sacrificing precision. The political keywords model uses the keywords ``lula", ``bolsonaro" and ``\#eleicoes2022". }
    \label{tab:results}
\end{table}

\subsection{Choice of Keywords}

We tested the PU-learning technique using many possible combinations of the hashtag \#eleicoes2022 and 10 keywords extracted\footnote{\label{kw}The complete list of keywords is: "\#eleicoes2022", "lula", "bolsonaro", "partido", "presidencia", "candidatura", "eleicoes", "eleitoral", "presidente", "debate","eleicao"} from the wikipedia page for the 2022 Brazilian elections\footnote{https://pt.wikipedia.org/wiki/Eleição\_presidencial\_no\_Brasil \_em\_2022}, as a way to discern the effect of that choice in the resulting classifier. Table \ref{tab:keyword_choice} shows the impact of the choice of keywords. It can be seen that increasing the number of keywords had two effects: increasing recall and reducing precision. These effects could be seen both in the classification using purely keywords as well as the two-step XGBoost trained on the data labeled by using these keywords.

\begin{table}[hb]
\small
\centering
\begin{tabular}{lrrrr} \toprule
\textbf{Model}          & \textbf{Accuracy} & \textbf{F1} & \textbf{Recall} & \textbf{Precision} \\\midrule
\textbf{Kw-based (3 kw)}     & 0.65              & 0.55        & 0.40            & 1.0                \\
\textbf{Kw-based (11 kw)}    & 0.70              & 0.63        & 0.49            & 0.98               \\
\bottomrule
\end{tabular}

\caption{Effect of the change in the number of keywords in the resulting models (keyword only or two-step PU XGBoost). In this context, "3 kw", for example, means that the first three keywords were used\textsuperscript{\ref{kw}}. Note how increasing number of keywords slightly changes recall and precision in the resulting models.}
\label{tab:keyword_choice}
\end{table}

A interesting insight from this table is that using the proposed method, it is possible to carefully choose keywords so as to better accomplish a desired result, for example, in an application where recall is desired, it might be interesting to increase the number of keywords, while the opposite can be said if precision is of interest. Since our goal was to be as precise as possible in the classification of comments to avoid detecting politicization where there was none, only 3 keywords were used in the creation of the final classifier. 

\textbf{Soccer as a control group.} To contrast the numbers and probability distributions, we conducted the same experiments with a different topic as the focus: soccer. We used the names of the 12 popular Brazilian teams as the seed keywords and ran the same PU Learning strategy with the same parameters in order to evaluate whether users shift topics to soccer with the same intensity as they do for politics.

\section{Characterizing Politicization}
\label{sec:results}
After building a classifier to categorize news posts and comments as political or non-political, we identify politicization by detecting topic shifts. Specifically, we look for comments classified as Brazilian politics, even when they are in response to non-political news articles.

\begin{figure}[ht]
    \centering

    \begin{subfigure}[t]{0.45\textwidth}    \centering
    \includegraphics[width=0.88\columnwidth]{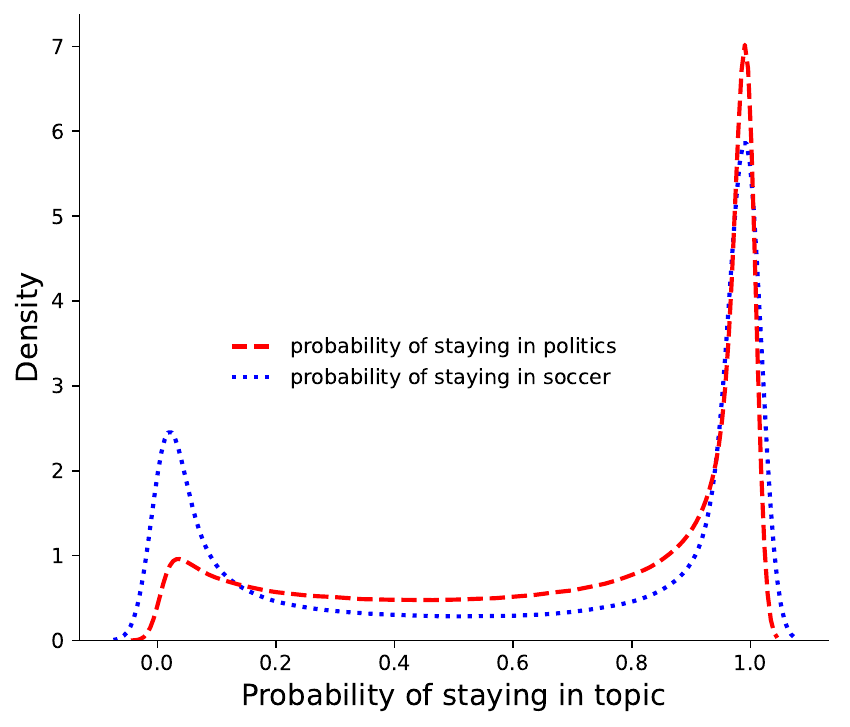}
    \caption{PDF of comment staying in topic.}
    \label{fig:p-transitions}
    \end{subfigure}
    
    \begin{subfigure}[t]{0.45\textwidth}    \centering
    \includegraphics[width=0.88\columnwidth]{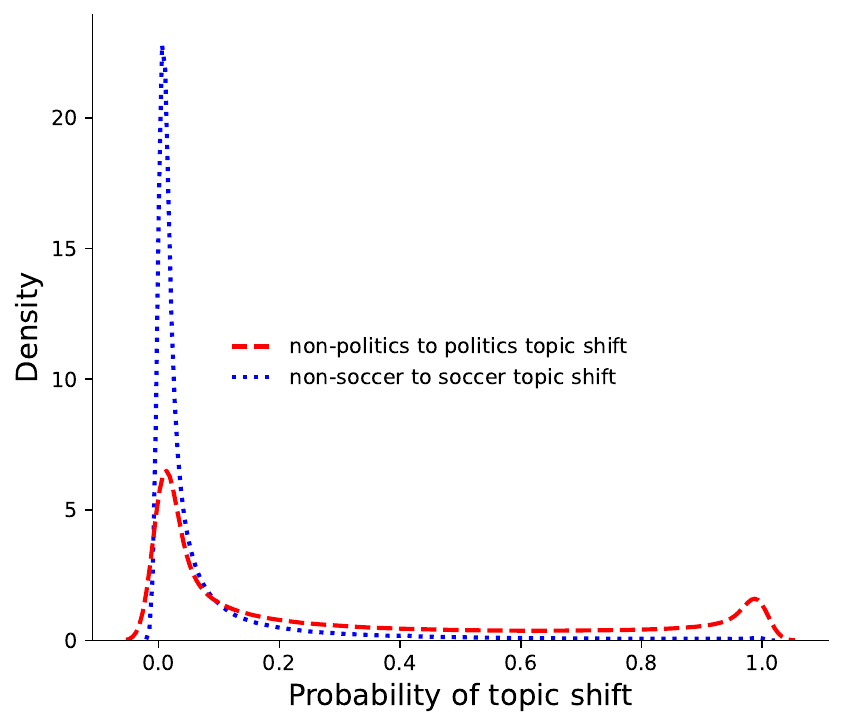}
    \caption{PDF of comment shifting the topic.}
    \label{fig:np-transitions}
    \end{subfigure}

    \caption{Probability density function (PDF) of topic shifts on all platforms. While discussions about politics and soccer typically remain on-topic, political comments frequently emerge in non-political discussions, causing a second peak in the probability of topic shifting in Figure~\ref{fig:np-transitions}.}
\end{figure}

As illustrated in Figure~\ref{fig:p-transitions}, it becomes evident that the dominant tendency is for comments to remain on the same topic as the original news article when it is either about politics or soccer. However, the most significant difference is depicted in Figure~\ref{fig:np-transitions}, which displays the probability density function of a comment shifting to politics/soccer when the original news article was unrelated. While only a few comments transition to soccer, with the majority of the distribution density centered around very low probabilities, the probability of a shift toward politics exhibits a more spread-out distribution along the x-axis, featuring a second peak around highly political comments. Specifically, while only 5\% of news articles shift toward soccer, a notable 35\% shift toward politics.

\textbf{Politicization by platform.} Tables~\ref{tab:politicalpu} and~\ref{tab:nonpoliticalpu} show the prevalence of political comments for news predicted as political and non-political, respectively. Note in Table~\ref{tab:politicalpu} that 94\%+ of all news posts triggered at least one political comment, what is consistent to what we expect. Also, the proportion of political comments in response to political news increased from 27, 15, and 31\% to 74, 64 and 70\% for YouTube, Twitter and TikTok when we compare with Table~\ref{tab:politicalratio1}, indicating that the classifier is indeed expanding the boundary of what political content looks like.

TikTok appears to be less politicized, with a higher percentage of non-political news, coupled with a low percentage of political comments on those news, while also having the lowest percentage of news with at least one political comment among all platforms. It is worth noting that YouTube and Twitter appear to be much more similar to one another than TikTok, as the same news sources posted much more political news to the first two social media platforms while prioritizing non-political posts on the latter.

The aforementioned differences and similarities between the studied platforms make it important to contrast the characteristics of topic shift on each of them, as their distinct features and public may influence this aspect.

\begin{figure}[t]
    \centering

    \begin{subfigure}[t]
    {0.45\textwidth}    \centering \includegraphics[width=0.85\columnwidth]{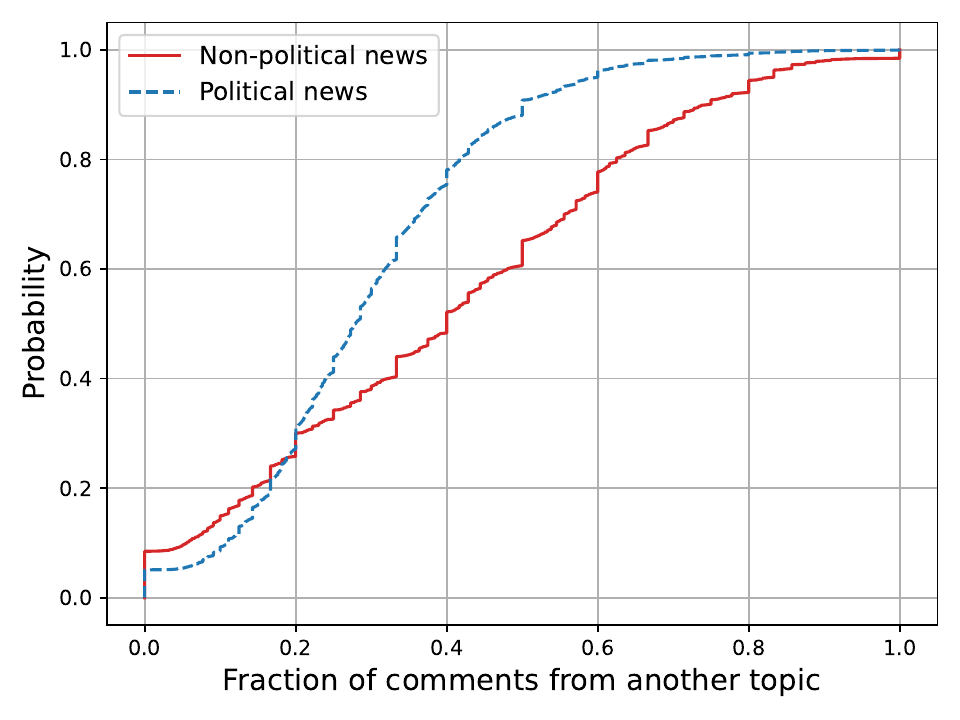}
    \caption{Twitter}
    \label{fig:cdf_twitter}
    \end{subfigure}
    
    \begin{subfigure}[t]{0.45\textwidth}    \centering
    \includegraphics[width=0.85\columnwidth]{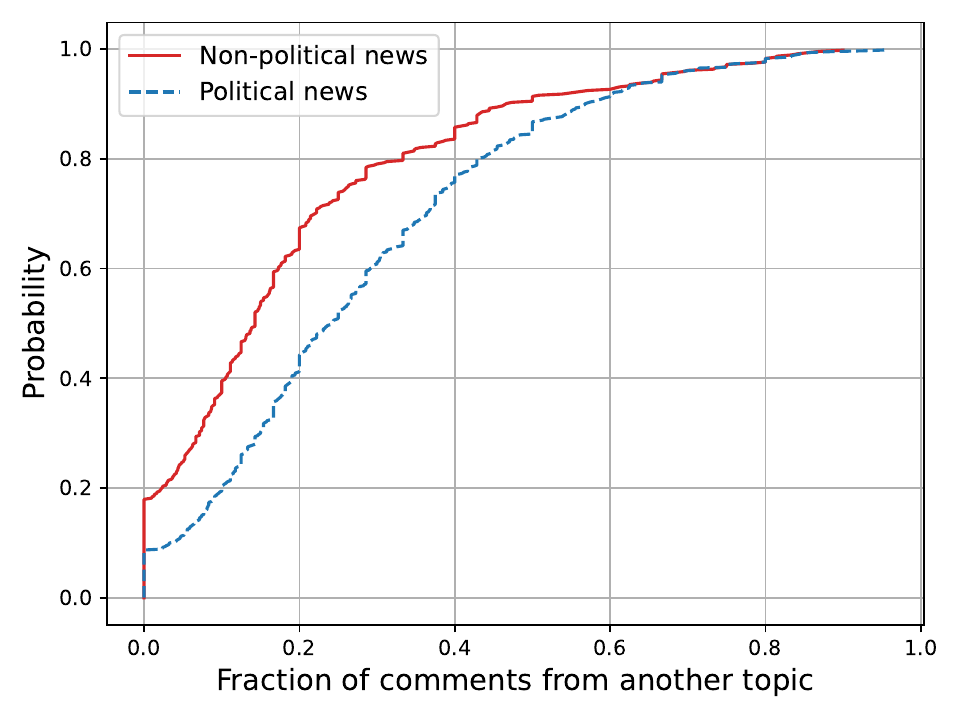}
    \caption{TikTok}
    \label{fig:cdf_tiktok}
    \end{subfigure}

    \caption{Cumulative Distribution Functions (CDFs) of the percentage of comments exhibiting Topic Shifts for Twitter and TikTok posts. Distribution for YouTube was omitted due to its similarity to the Twitter one.}
\end{figure}

Figures \ref{fig:cdf_twitter} and \ref{fig:cdf_tiktok} show the Cumulative Distribution Function of Topic Shifts on Twitter and TikTok comment sections, respectively. The distribution for YouTube was omitted due to it showing a very similar pattern to the Twitter one, despite being less politicized than the latter.

To allow for a more accurate comparison, the data used in these distributions was filtered to include only the dates when data was available for all three platforms (2022-08-26 to 2022-11-01). However, the analyses were also performed on the entire dataset, yielding similar results.


\begin{table}[t]
\small
\centering
\begin{tabular}{cccccc}
\toprule
\multicolumn{1}{c}{} & \multicolumn{1}{c}{} & \multicolumn{3}{c}{\textbf{Comments in P News Posts}} \\
\cmidrule(rl){3-5}
\textbf{Platform} & \textbf{P Posts} & {P} & {Non-P} & {At least one P} \\
\midrule
YouTube & 47\% & 74\% & 26\% & 97\% \\
Twitter & 45\% & 64\% & 36\% & 94\% \\
TikTok & 30\% & 70\% & 30\% & 94\% \\
\bottomrule
\end{tabular}

\caption{Ratio of political (P) posts per platform predicted by
the PU Learning-based classifier.}
\label{tab:politicalpu}
\end{table}


\begin{table}[t]
\centering
\small
\begin{tabular}{cccccc}
\toprule
\multicolumn{1}{c}{} & \multicolumn{1}{c}{} & \multicolumn{3}{c}{\textbf{Comments in Non-P News Posts}} \\
\cmidrule(rl){3-5}
\textbf{Platform} & \textbf{Non-P  Posts} & {P} & {Non-P} & {At least one P} \\
\midrule
YouTube & 53\% & 25\% & 75\% & 78\% \\
Twitter & 55\% & 35\% & 65\% & 61\% \\
TikTok & 70\% & 24\% & 76\% & 60\% \\
\bottomrule
\end{tabular}
\caption{Ratio of non-political (non-P) posts per platform predicted by the PU Learning-based classifier.}
\label{tab:nonpoliticalpu}
\end{table}

These distributions show that topic shift is a phenomenon that is more prevalent in non-political news, leading the conversation to political topics. In fact, on both YouTube and Twitter, the non-political news CDF crosses the political news CDF on topic shift$\approx$20\% and probability $\le$ 50\%. This means that more than 50\% of posts in non-political news, show more topic shift than their political counterparts. TikTok, on the other hand, does not follow this pattern, with non-political and political news having similar topic shift distributions. This variation on TikTok could be explained by several factors, including but not limited to: 

\begin{itemize}
    \item TikTok's short video format, which may inhibit deep or serious discussions;
    \item TikTok's younger audience \cite{kanthawala2022s}, which may be less interested in politics.
\end{itemize}


\subsection{Finding Most Politicized Topics}

\begin{table*}[ht]
\small
        \centering
\begin{tabular}{llrl} \toprule
\textbf{Topic}      & \textbf{Representative Words} & \makecell[r]{\textbf{Political}\\\textbf{Comments}} & \textbf{Sample}                                                                                    \\\midrule

\textbf{Soccer} &  Sports, soccer, Atlético         & 5\%                                          & \multirow{2}{*}{\makecell[l]{\textbf{News.}``PALMEIRAS 1 X 1 FLAMENGO $|$ Best Moments $|$\\ Brasileirão 2022 round 23."\\\textbf{Politicized Comment.}``Bolsonaro made Brasil worse."\textsuperscript{\getrefnumber{translate}}}} \\
\multicolumn{2}{l}{\cite{politicization_world_cup}}  & &
\\ [15pt]

\textbf{NBA} & Basketball, Knicks, Lakers             & 9\%                                            &\multirow{2}{*}{\makecell[l]{\textbf{News.}``Video shows Draymond Green punching Jordan Poole during \\Golden State Warriors practice."\\\textbf{Politicized Comment.}``And how's the former prisioner?"(Lula)\textsuperscript{\getrefnumber{translate}} }} \\ 
\multicolumn{2}{l}{\cite{zhao2022navigating}}  & &
\\ [15pt]

\textbf{AI}  & Chatbot, chatgpt, robot  & 13\%                                            & \multirow{2}{*}{\makecell[l]{\textbf{News.}``Scientists develop new 'Terminator' robot"\\\textbf{Politicized Comment.} ``This one is nothing, we already have the \\nine-fingered mollusk"(Lula)\textsuperscript{\getrefnumber{translate}}}} \\ 
\multicolumn{2}{l}{\cite{ossewaarde2020national}}  & &
\\ [15pt]

\textbf{NFL}    & Quarterback, football, 49ers & 14\%                                           & \multirow{2}{*}{\makecell[l]{\textbf{News.}``MISSSSSSSS Matt Prater misses NFL's 50-yard field goal, \\Patriots vs. Cardinals still tied after Ari Aguiar's 'hex' \#ESPNnoStarPlus"\\\textbf{Politicized Comment.}``We are suffering, Bolsonaro's help in Brasilia is missing."\textsuperscript{\getrefnumber{translate}} }}\\ 
\multicolumn{2}{l}{\cite{dagnes2019us}}  & &
\\ [20pt]

\textbf{Pets}      & Dogs, pets, puppy   & 19\%                                            & \multirow{2}{*}{\makecell[l]{\textbf{News.}``How well can you decipher monkey language?"\\\textbf{Politicized Comment.}``L with his pet judge."(Lula)\textsuperscript{\getrefnumber{translate}}}}\\ 
\multicolumn{2}{l}{\cite{digard2004construction}}  &  &
\\ \bottomrule
\end{tabular}

\caption{Least politicized non-political topics across all platforms. Note how most topics relate to soft news. For each topic, we reference recent work that studied the topic through the lens of politicization.}

\label{tab:least_politicized}
\end{table*}

\begin{table*}[ht]
\small
        \centering
\begin{tabular}{llrll} \toprule
\textbf{Topic}      & \textbf{Representative Words} & \makecell[r]{\textbf{Political}\\\textbf{Comments}} & \textbf{Sample}                                                                                    \\\midrule

\textbf{Economy}    & Inflation, economy, recession & 54\%                                            & \multirow{2}{*}{\makecell[l]{\textbf{News.}``Ceasa in Rio de Janeiro catches fire; warehouses are looted."\\\textbf{Politicized Comment.} ``People are already blaming Lula  and he is \\not even president yet."\textsuperscript{\getrefnumber{translate}}}}\\ 
\multicolumn{2}{l}{\cite{schaffner2016misinformation}}  & &
\\ [15pt]

\textbf{Fossil Fuel} & Prices, fuel, oil             & 49\%                                            &\multirow{2}{*}{\makecell[l]{\textbf{News.}``How the decline in oil affects Petrobras, who plans to increase\\ production of the fossil fuel."\\\textbf{Politicized Comment.} ``If it is up to Paulo Guedes and Lula,\\  Petrobrás will be sold. Only Ciro Gomes can save it."\textsuperscript{\getrefnumber{translate}}}}       \\ 
\multicolumn{2}{l}{\cite{healy2017politicizing}}  & &
\\ [25pt]

\textbf{Crime}      & Suspect, police, criminals         & 49\%                                            & \multirow{2}{*}{\makecell[l]{\textbf{News.}``PRF rescued the children and said that the father could\\ be charged with abandoning a child \#g1"\\\textbf{Politicized Comment.}``Bolsonarists are Nazis, homophobes, \\antidemocratic..."\textsuperscript{\getrefnumber{translate}}}} \\ 
\multicolumn{2}{l}{\cite{daxecker2016politicization}}  & &
\\ [25pt]

\textbf{Drugs} &  Drugs, cocaine, smuggling   & 45\%                                            & \multirow{2}{*}{\makecell[l]{\textbf{News.}``PF arrests man with almost 1 ton of marijuana on \\Via Dutra, in Rio"\\\textbf{Politicized Comment.}``...Do people arrested for drug dealing \\vote for Bolsonaro?"\textsuperscript{\getrefnumber{translate}}}} \\ 
\multicolumn{2}{l}{\cite{brown2022social}}  & &
\\ [25pt]

\textbf{Banking}  & Bank, debt, investiment  & 45\%                                            & \multirow{2}{*}{\makecell[l]{\textbf{News.}``Market: Banks closed on Nossa Senhora da Aparecida holiday, \\but reopen on Thursday"\\\textbf{Politicized Comment.}``I voted for Lula on the first round, but he \\ talks to the devil so I will vote for Bolsonaro on the second round"\textsuperscript{\getrefnumber{translate}}}}\\ 
\multicolumn{2}{l}{\cite{gonzalez2021financialization}}  & &
\\ [15pt] \\\bottomrule
\end{tabular}

\caption{Most politicized non-political topics across all platforms, excluding misclassifications. Most topics relate to hard news. For each topic, we reference recent work that studied the topic through the lens of politicization.}
\label{tab:most_politicized}
\end{table*}

 The classifier described in the previous sections is able to predict when a piece of news or comment is political. This is enough to ascertain whether politicization happens or not in the context of online Brazilian news' comment sections. Now, we seek to find out which topics are more subject to politicization. We identify topics using BERTopic~\cite{grootendorst2022bertopic}, a topic modeling technique based on BERT embeddings and c-TF-IDF that produces interpretable topics.

The news posts were split into political or non-political based on the classifier output, and topics were produced for each of those two groups, with each topic covering at least 100 news posts. After assigning each news post to a topic, we calculate the percentage of comments for each topic whose classification was different from the content it referred to. We can then identify which topics were more likely to be shifted towards or away from politics, and since the topics are highly interpretable, we can manually spot misclassified news and exclude them from the analysis.


When looking at the topics generated by BERTopic, we can identify a variety of relevant events that happened in 2022/2023. On non-political news, we identify 52 topics, with examples such as soccer, cryptocurrencies, the NFL, and even chatbots. Meanwhile, on political news, we identify 19 topics, including elections, corruption (in a variety of areas), and candidate debates.

Before discussing the most politicized topics, it is important to acknowledge some possibly misclassified topics. On the non-political topics, we see voter registration card (information about how to get the document), trucker road blockades (when the focus is on the events and not politics) and daily news (which most of the time contain political segments that may lead to political comments), while on the political side we see Russian, Chinese and Latin American Governments, possibly due to overlapping vocabulary between these topics and news about Brazilian politics.

That said, excluding these possibly misclassified topics, the most politicized topics included the economy, fossil fuel, crime, and drugs. Meanwhile, the least politicized topics related to entertainment and lifestyle, with topics such as sports, pets, food, and celebrities. Tables \ref{tab:least_politicized} and \ref{tab:most_politicized} show, respectively, the top-5 least and most politicized non-political topics. It is possible to see an extreme difference in the percentage of political comments, with the least politicized having less than 20\% of comments classified as political, while the most politicized have around 50\% political comments. Notice that, although all topics show politicized comments, a fair amount of nitpicking was required to find politicization in Table \ref{tab:least_politicized}, with many of the videos not having even a single political comment, while in Table \ref{tab:most_politicized} sampling a single random video and 2 or so comments classified as political was enough to find very explicit examples of politicization.


Comparing non-political with political topics, it can be seen that even the least politicized political topics have a percentage of political comments comparable to the most politicized non-political topics. In fact, only news related to the Chinese government were less politicized, on average, than the most politicized non-political topic, with 36\% political comments. All other topics have a greater percentage of political comments. This finding suggests that, while the classifier is not perfect, it is able to give higher probabilities in general for comments in political news. This is evidenced by the fact that non-political news have 26\% of political comments, even when considering misclassified topics, while  political news have 76\% political comments. 

Interestingly, some of the most politicized topics include religion in politics, elections, debates, and protests (focusing on politicians' reactions), which were very relevant topics in the context of the 2022 Brazilian elections, with candidates, such as Father Kelmon and many others, appealing to voters through religion. Moreover, some topics that previous studies defined as politicized, but aren't political in a vacuum had a significant percentage of political comments ($>$30\%).  Examples of this include vaccination and climate change, with some news that might seem innocuous, such as ``COP-27: Why Greta Thunberg is avoiding the UN climate conference this year", being extremely politicized in the comments, in part due to Bolsonaro's previous clash with Greta. Surprisingly, the topic of the death of Queen Elizabeth II was also considerably shifted towards Brazilian Politics.

\subsection{Temporal Changes in Topic Shifts}

\begin{figure}[ht]
    \centering
    \includegraphics[width=1.1\columnwidth]{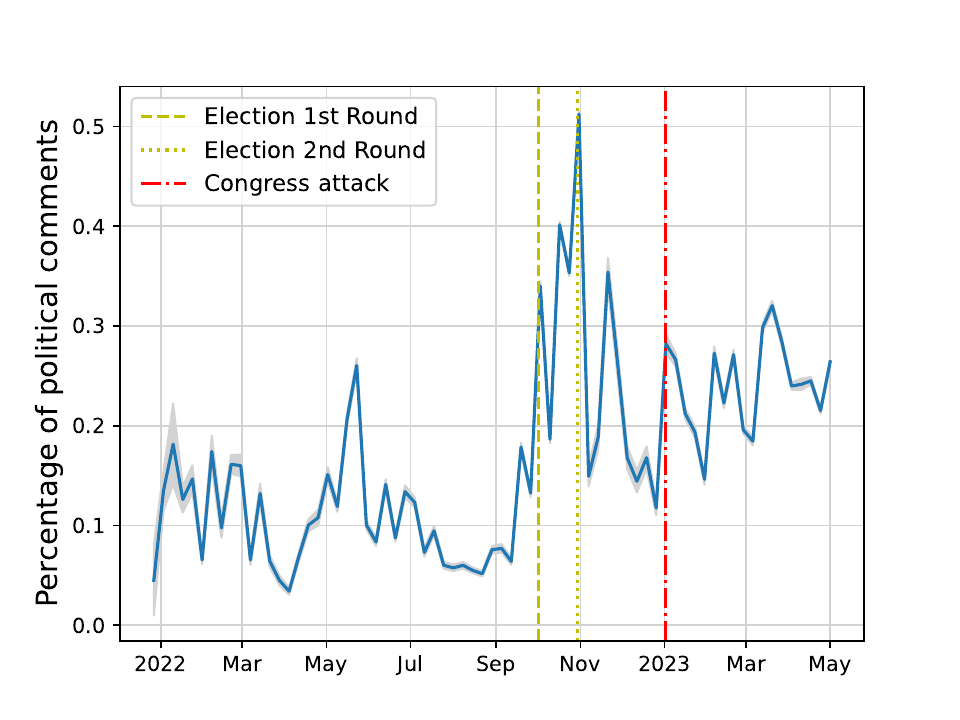}
    \caption{Ratio of YouTube comments which are a topic shift from the news posts, by week. During the two rounds of the Brazilian elections, we see spikes on non-political content being politicized. The gray area is the confidence interval.}
    \label{fig:dates}
\end{figure}

Exploring the temporal dynamics of topic shifts is also important for better understanding how this metric relates to societal politicization. This section focuses on YouTube because it is the platform for which we have the longest data collection period. Figure \ref{fig:dates} shows the frequency of topic shifts in each week's comments on videos classified as \emph{non-political}. Some significant events concerning Brazilian politics are also highlighted. 

We see a increase in topic shifts towards politics in the week of the elections. The ratio of political comments continues to grow until ultimately peaking during the week of the second round of elections, possibly due to the discussions between each candidate's supporters. The percentage of political comments starts to reduce after Lula's victory, before showing another peek in the week of the Brazilian congress attack/Lula's inauguration as president, after which the rate of politic comments remained high for the remainder of the studied period.

 High rates of political comments in a given date are a good indication, but do not necessarily translate to politicization of a given topic. As seen in table \ref{tab:most_politicized} some topics are on average more politicized than others considering a snapshot. We propose that when a topic is being politicized, it takes less comments until a political comment can be seen in response to a given news piece. Over time, this means that a topic that was politicized during the elections would have its topic shifted faster after this period.

\begin{figure}[ht]
    \centering
    \includegraphics[width=1.1\columnwidth]{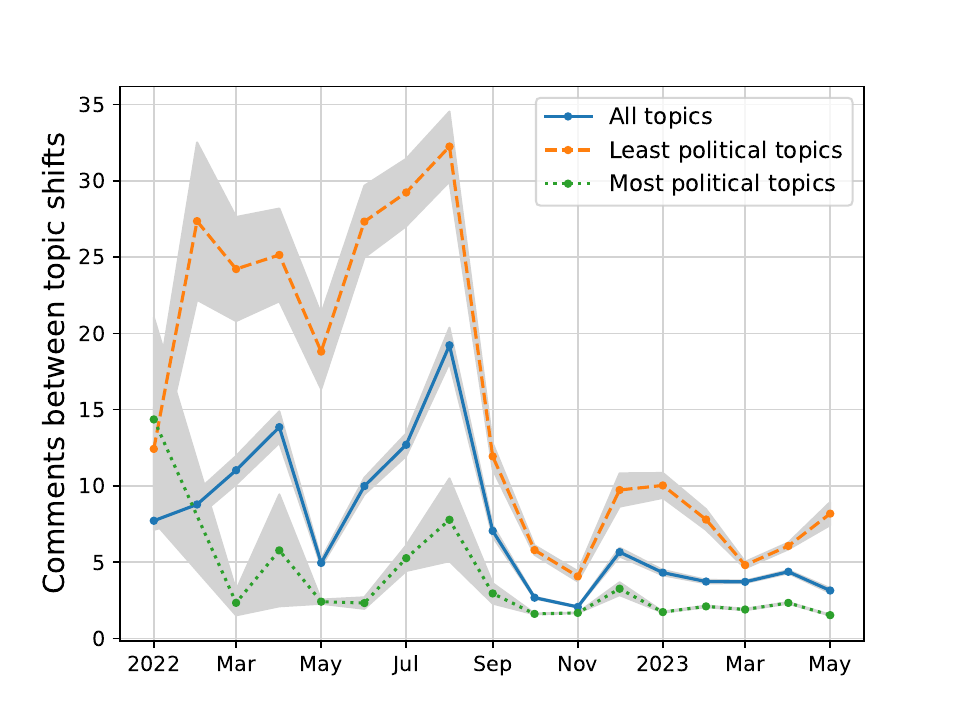}
    \caption{Average number of comments between topic shifts toward politics for all YouTube videos by month. A smaller number means that users shift more frequently toward politics. In the start of 2022 there were few videos of the most politicized topics, hence the larger confidence interval (gray area). Again, an increase in topic shifts toward politics coincides with the election period.}
    \label{fig:speed_of_shifts}
\end{figure}

Figure \ref{fig:speed_of_shifts} shows the mean number of comments before a topic shift occurs on a given month on non-political news, separated in three groups: the top-5 most politicized topics, the top-5 least politicized topics (tables \ref{tab:least_politicized} and \ref{tab:most_politicized}) and all topics. Looking at the least politicized topics it is possible to see that the tendencies of politicization align (although not always perfectly) to those seen in figure \ref{fig:dates}, seeing that as the ratio of political comments grow, the speed of which commenters change the news topic increases. In fact, in the lead up to the elections, the number of comments before a topic shift decreased drastically, remaining relatively low for the rest of the studied period, which could be an indication that politicization took place. The same could be said for the entire dataset, with similar tendencies taking place.

However, when looking at the most politicized non-political topics, they seem to already have been politicized even before the elections and other relevant political events in Brazil took place. Many of those (e.g. drugs~\cite{brown2022social}) were found to be politicized by other researchers in other contexts.

These findings suggest that, for some topics, especially those that are least political, there was some degree of politicization on the lead up to and after the 2022 Brazilian elections, which could have been influenced by the political events that took place during this period.

Another interesting aspect to consider is whether or not the occurrence of a topic shift on the comment sections of a given news piece led to more frequent shifts in subsequent comments. To answer this question, we looked through how many comments were posted between each topic shift for each piece of news. However, the ratio of topic shifts does not seem to increase nor decrease after a topic shifting comment is posted as there is no statistically significant difference between the average number of comments before the first topic shift and between each other topic shift. This could suggest that generally commenters post independently from each other.

\section{Conclusions and Future Work}
\label{sec:conclusions}

We study topic shifts over social media conversations as a novel strategy to measure politicization, more specifically in the context of the 2022 Brazilian presidential elections. While politicization is often studied on specific topics or mentioned in a cursory way, we propose a computational method that directly observes and quantifies politicization using \emph{topic shifts}, i.e., the change of a topic by social media users participating in a discussion. 

Starting from a few political keywords that work as seeds, we conducted a two-step PU Learning strategy that learns the boundary between political and non-political content. PU is an effective method to study politicization given that it demands only a few positive keywords to serve as seeds, it does not require the labeling of non-political content, which is inherently diverse and heterogeneous in nature, and it has only a few calibration parameters which are not very sensitive according to both the literature and our experiments. We evaluated the results against an annotated dataset, and our method achieves around 92\% and 80\% F1 scores on news posts and comments, respectively.



Our results indicate that, indeed, politicization is a prevalent social process in social media, aligned with previous research on Reddit communities~\cite{rajadesingan2021political}. We found that
topics previously found to be politicized in other contexts,
for example vaccination and climate change, tended to have
a high percentage of political comments. We also found that even the least political topics (e.g., Sports and pets) had an increased rate of topic shifts towards politics during and after the elections, suggesting that politicization took place.
We believe our work solidifies a recent trend that, since political talk may occur anywhere~\cite{rajadesingan2021political}, looking for behavioral patterns when topics drift and merge gives us the opportunity to contrast behaviors, build null models, and compare the observed political behavior with that of control groups. For example, our results reinforce how motivated reasoning -- the influence of our motivations and goals in our reasoning -- is a cognitive process that is highly tied to politicization~\cite{bolsen2018partisanship}.

In future work, we plan to better link politicization with polarization and attempt to establish potential correlations and cause-and-effect connections between those two core political processes. We will also be looking at user profiles; are there a few users that politicize everything?

\section{Broader Perspective, Ethics and Competing Interests}
All data we use from TikTok, Twitter, and YouTube was publicly available when we collected it. Additionally, all labels were created by people directly involved in the research project. To avoid compromising individual users, any comment quoted on this paper was translated, paraphrased, and modified (while keeping the general meaning). 

Our work focuses on assessing and characterizing politicization without using any manual labels, which can accelerate and encourage further research in the political sciences. While we acknowledge that the accuracy of the classifier in the range of 84\% is a potential threat to the validity of the results, since a correct topic shift is a result of a correct classification of both the news posts and the comment, we believe the effect of the prediction error is minimized due to two efforts: (1) we manually discarded the misclassified topical clusters of news posts, and (2) since each post receives on average tenths or hundreds of comments (Table~\ref{tab:statistics}), the errors tend to cancel out and a signal of politicization still emerges, as the analysis in Tables~\ref{tab:least_politicized} and \ref{tab:most_politicized}, and Figure~\ref{fig:speed_of_shifts} made clear.

\section{Acknowledgments}

The authors would like to thank Igor Joaquim da Silva Costa, Mateus F. Zaparoli, Thaís F. da Silva, Victor B. Thomé and Davi Fraga from UFMG, and Arthur R. S. da Costa from Unicamp for their valuable contributions and feedback during the research.

This work is partially supported by CNPq, CAPES, FAPEMIG and FAPESP, and by the projects INCT-Cyber, CIIA-Saúde, and PCA/MPMG.

The authors declare no conflicts of interest.

\bibliography{bibliography}

\begin{thebibliography}{58}
\providecommand{\natexlab}[1]{#1}

\bibitem[{Baum and Groeling(2008)}]{new_media_bias}
Baum, M.~A.; and Groeling, T. 2008.
\newblock New Media and the Polarization of American Political Discourse.
\newblock \emph{Political Communication}, 25(4): 345--365.

\bibitem[{Bay(2018)}]{bay2018weaponizing}
Bay, M. 2018.
\newblock Weaponizing the haters: The Last Jedi and the strategic politicization of pop culture through social media manipulation.
\newblock \emph{First Monday}.

\bibitem[{Bekker and Davis(2020)}]{learning_from_positive_unlabeled}
Bekker, J.; and Davis, J. 2020.
\newblock Learning from Positive and Unlabeled Data: A Survey.
\newblock \emph{Mach. Learn.}, 109(4): 719–760.

\bibitem[{Bolsen and Druckman(2015)}]{politicization_science}
Bolsen, T.; and Druckman, J.~N. 2015.
\newblock {Counteracting the Politicization of Science}.
\newblock \emph{Journal of Communication}, 65(5): 745--769.

\bibitem[{Bolsen and Druckman(2018)}]{bolsen2018partisanship}
Bolsen, T.; and Druckman, J.~N. 2018.
\newblock Do partisanship and politicization undermine the impact of a scientific consensus message about climate change?
\newblock \emph{Group Processes \& Intergroup Relations}, 21(3): 389--402.

\bibitem[{Boynton and Richardson~Jr(2016)}]{boynton2016agenda}
Boynton, G.; and Richardson~Jr, G.~W. 2016.
\newblock Agenda setting in the twenty-first century.
\newblock \emph{New Media \& Society}, 18(9): 1916--1934.

\bibitem[{Brown and Midberry(2022)}]{brown2022social}
Brown, D.~K.; and Midberry, J. 2022.
\newblock Social media news production, emotional Facebook reactions, and the politicization of drug addiction.
\newblock \emph{Health communication}, 37(3): 375--383.

\bibitem[{Brummette et~al.(2018)Brummette, DiStaso, Vafeiadis, and Messner}]{fake_news_policitization}
Brummette, J.; DiStaso, M.; Vafeiadis, M.; and Messner, M. 2018.
\newblock Read all about it: The politicization of “fake news” on Twitter.
\newblock \emph{Journalism \& Mass Communication Quarterly}, 95(2): 497--517.

\bibitem[{Chen and Guestrin(2016)}]{Chen:2016:XST:2939672.2939785}
Chen, T.; and Guestrin, C. 2016.
\newblock {XGBoost}: A Scalable Tree Boosting System.
\newblock In \emph{Proc. of the 22nd ACM SIGKDD Int'l Conf. on Knowledge Discovery and Data Mining}, KDD '16, 785--794. New York, NY, USA: ACM.

\bibitem[{Chinn, Hart, and Soroka(2020)}]{politicization_polarization}
Chinn, S.; Hart, P.~S.; and Soroka, S. 2020.
\newblock Politicization and Polarization in Climate Change News Content, 1985-2017.
\newblock \emph{Science Communication}, 42(1): 112--129.

\bibitem[{Dagnes and Dagnes(2019)}]{dagnes2019us}
Dagnes, A.; and Dagnes, A. 2019.
\newblock Us vs. them: Political polarization and the politicization of everything.
\newblock \emph{Super Mad at Everything All the Time: Political Media and Our National Anger}, 119--165.

\bibitem[{Daxecker and Prins(2016)}]{daxecker2016politicization}
Daxecker, U.~E.; and Prins, B.~C. 2016.
\newblock The politicization of crime: electoral competition and the supply of maritime piracy in Indonesia.
\newblock \emph{Public choice}, 169(3-4): 375--393.

\bibitem[{Diaz et~al.(2022)Diaz, Hanna, Hughes, Lehmann, and Medford}]{Politicization_Ivermectin}
Diaz, M.~I.; Hanna, J.~J.; Hughes, A.~E.; Lehmann, C.~U.; and Medford, R.~J. 2022.
\newblock {The Politicization of Ivermectin Tweets During the COVID-19 Pandemic}.
\newblock \emph{Open Forum Infectious Diseases}, 9(7).
\newblock Ofac263.

\bibitem[{Digard(2004)}]{digard2004construction}
Digard, J.-P. 2004.
\newblock La construction sociale d’un animal domestique: le pitbull.
\newblock \emph{Anthropozoologica}, 39(1): 17--26.

\bibitem[{Edelman et~al.(2020)Edelman, Wolff, Montagne, and Bail}]{edelman_computational_2020}
Edelman, A.; Wolff, T.; Montagne, D.; and Bail, C.~A. 2020.
\newblock Computational {Social} {Science}.
\newblock \emph{Annual Review of Sociology}, 46.
\newblock 00000.

\bibitem[{Elkan and Noto(2008)}]{elkan2008learning}
Elkan, C.; and Noto, K. 2008.
\newblock Learning classifiers from only positive and unlabeled data.
\newblock In \emph{Proc. of the 14th ACM SIGKDD int'l conf. on Knowledge discovery and data mining}, 213--220.

\bibitem[{Fernandes et~al.(2020)Fernandes, Ademir~de Oliveira, Motta~de Campos, and Gomes}]{polarization_Brazilian_Elections}
Fernandes, C.~M.; Ademir~de Oliveira, L.; Motta~de Campos, M.; and Gomes, V.~B. 2020.
\newblock Political polarization in the Brazilian Election Campaign for the Presidency of Brazil in 2018: an analysis of the social network Instagram.
\newblock \emph{Int'l J. Soc. Sci. Stud.}, 8: 119.

\bibitem[{Fusilier et~al.(2015)Fusilier, Montes-y G{\'o}mez, Rosso, and Cabrera}]{fusilier2015detecting}
Fusilier, D.~H.; Montes-y G{\'o}mez, M.; Rosso, P.; and Cabrera, R.~G. 2015.
\newblock Detecting positive and negative deceptive opinions using PU-learning.
\newblock \emph{Information processing \& management}, 51(4): 433--443.

\bibitem[{Gonz{\'a}lez-L{\'o}pez(2021)}]{gonzalez2021financialization}
Gonz{\'a}lez-L{\'o}pez, F. 2021.
\newblock The financialization of social policy and the politicization of student debt in Chile.
\newblock \emph{Journal of Cultural Economy}, 14(2): 176--193.

\bibitem[{Graham, Avery, and Park(2015)}]{social_media_communications}
Graham, M.~W.; Avery, E.~J.; and Park, S. 2015.
\newblock The role of social media in local government crisis communications.
\newblock \emph{Public Relations Review}, 41(3): 386--394.

\bibitem[{Grootendorst(2022)}]{grootendorst2022bertopic}
Grootendorst, M. 2022.
\newblock BERTopic: Neural topic modeling with a class-based TF-IDF procedure.
\newblock \emph{arXiv preprint arXiv:2203.05794}.

\bibitem[{Grover et~al.(2019)Grover, Kar, Dwivedi, and Janssen}]{GROVER2019438}
Grover, P.; Kar, A.~K.; Dwivedi, Y.~K.; and Janssen, M. 2019.
\newblock Polarization and acculturation in US Election 2016 outcomes – Can twitter analytics predict changes in voting preferences.
\newblock \emph{Technological Forecasting and Social Change}, 145: 438--460.

\bibitem[{Hart, Chinn, and Soroka(2020)}]{Politicization_and_Polarization_COVID}
Hart, P.~S.; Chinn, S.; and Soroka, S. 2020.
\newblock Politicization and Polarization in COVID-19 News Coverage.
\newblock \emph{Science Communication}, 42(5): 679--697.

\bibitem[{Healy and Barry(2017)}]{healy2017politicizing}
Healy, N.; and Barry, J. 2017.
\newblock Politicizing energy justice and energy system transitions: Fossil fuel divestment and a “just transition”.
\newblock \emph{Energy policy}, 108: 451--459.

\bibitem[{Holmberg(2015)}]{1ow_carb_diet}
Holmberg, C. 2015.
\newblock Politicization of the Low-Carb High-Fat Diet in Sweden, Promoted On Social Media by Non-Conventional Experts.
\newblock \emph{Int. J. E-Polit.}, 6(3): 27–42.

\bibitem[{Howison, Crowston, and Wiggins(2011)}]{validity}
Howison, J.; Crowston, K.; and Wiggins, A. 2011.
\newblock Validity issues in the use of social network analysis with digital trace data.
\newblock \emph{Journal of the Assoc. for Information Systems}, 12.

\bibitem[{Kane and Luo(2018)}]{communities_shape_political_beliefs}
Kane, B.; and Luo, J. 2018.
\newblock Do the Communities We Choose Shape our Political Beliefs? A Study of the Politicization of Topics in Online Social Groups.
\newblock In \emph{2018 IEEE International Conference on Big Data (Big Data)}, 3665--3671.

\bibitem[{Kanthawala et~al.(2022)Kanthawala, Cotter, Foyle, and DeCook}]{kanthawala2022s}
Kanthawala, S.; Cotter, K.; Foyle, K.; and DeCook, J.~R. 2022.
\newblock It's the Methodology For Me: A Systematic Review of Early Approaches to Studying TikTok.
\newblock In \emph{HICSS}, 1--17.

\bibitem[{Karimi et~al.(2021)Karimi, Tang, Weiss, and Huang}]{teacher_pu_learning}
Karimi, H.; Tang, J.; Weiss, X.; and Huang, J. 2021.
\newblock Automatic Identification of Teachers in Social Media using Positive Unlabeled Learning.
\newblock In \emph{2021 IEEE International Conference on Big Data (Big Data)}, 643--652.

\bibitem[{Landis and Koch(1977)}]{KochLandis77}
Landis, J.~R.; and Koch, G.~G. 1977.
\newblock The Measurement of Observer Agreement for Categorical Data.
\newblock \emph{Biometrics}, 33(1): 159--174.

\bibitem[{Layton et~al.(2021)Layton, Smith, Moseley, and Cohen}]{layton2021demographic}
Layton, M.~L.; Smith, A.~E.; Moseley, M.~W.; and Cohen, M.~J. 2021.
\newblock Demographic polarization and the rise of the far right: Brazil’s 2018 presidential election.
\newblock \emph{Research \& Politics}, 8(1): 2053168021990204.

\bibitem[{Lazer et~al.(2009)Lazer, Pentland, Adamic, Aral, Barabási, Brewer, Christakis, Contractor, Fowler, Gutmann, Jebara, King, Macy, Roy, and Alstyne}]{computational_social_science}
Lazer, D.; Pentland, A.; Adamic, L.; Aral, S.; Barabási, A.-L.; Brewer, D.; Christakis, N.; Contractor, N.; Fowler, J.; Gutmann, M.; Jebara, T.; King, G.; Macy, M.; Roy, D.; and Alstyne, M.~V. 2009.
\newblock Computational Social Science.
\newblock \emph{Science}, 323(5915): 721--723.

\bibitem[{Li et~al.(2014)Li, Chen, Liu, Wei, and Shao}]{li2014spotting}
Li, H.; Chen, Z.; Liu, B.; Wei, X.; and Shao, J. 2014.
\newblock Spotting fake reviews via collective positive-unlabeled learning.
\newblock In \emph{2014 IEEE int'l conf. on data mining}, 899--904. IEEE.

\bibitem[{Ling et~al.(2022)Ling, Blackburn, De~Cristofaro, and Stringhini}]{ling2022slapping}
Ling, C.; Blackburn, J.; De~Cristofaro, E.; and Stringhini, G. 2022.
\newblock Slapping Cats, Bopping Heads, and Oreo Shakes: Understanding Indicators of Virality in TikTok Short Videos.
\newblock In \emph{14th ACM Web Science Conference 2022}, 164--173.

\bibitem[{Liu(2007)}]{liu_web_data_mining}
Liu, B. 2007.
\newblock \emph{Web Data Mining: Exploring Hyperlinks, Contents, and Usage Data}.
\newblock Data-Centric Systems and Applications. Springer.
\newblock ISBN 978-3-540-37882-2.

\bibitem[{Liu et~al.(2002)Liu, Lee, Yu, and Li}]{liu2002partially}
Liu, B.; Lee, W.~S.; Yu, P.~S.; and Li, X. 2002.
\newblock Partially supervised classification of text documents.
\newblock In \emph{ICML}, volume~2, 387--394. Sydney, NSW.

\bibitem[{Liu and Wu(2020)}]{fake_news_early_detection}
Liu, Y.; and Wu, Y.-F.~B. 2020.
\newblock FNED: A Deep Network for Fake News Early Detection on Social Media.
\newblock \emph{ACM Trans. Inf. Syst.}, 38(3).

\bibitem[{Medina~Serrano, Papakyriakopoulos, and Hegelich(2020)}]{medina2020dancing}
Medina~Serrano, J.~C.; Papakyriakopoulos, O.; and Hegelich, S. 2020.
\newblock Dancing to the partisan beat: A first analysis of political communication on TikTok.
\newblock In \emph{12th ACM conference on web science}, 257--266.

\bibitem[{Meier et~al.(2021)Meier, Mutz, Glathe, Jetzke, and H{\"o}lzen}]{politicization_world_cup}
Meier, H.~E.; Mutz, M.; Glathe, J.; Jetzke, M.; and H{\"o}lzen, M. 2021.
\newblock Politicization of a contested mega event: The 2018 FIFA World Cup on Twitter.
\newblock \emph{Communication \& Sport}, 9(5): 785--810.

\bibitem[{Montag, Yang, and Elhai(2021)}]{montag2021psychology}
Montag, C.; Yang, H.; and Elhai, J.~D. 2021.
\newblock On the psychology of TikTok use: A first glimpse from empirical findings.
\newblock \emph{Frontiers in public health}, 9: 641673.

\bibitem[{Oschatz, Stier, and Maier(2022)}]{political_news_coverage}
Oschatz, C.; Stier, S.; and Maier, J. 2022.
\newblock Twitter in the News: An Analysis of Embedded Tweets in Political News Coverage.
\newblock \emph{Digital Journalism}, 10(9): 1526--1545.

\bibitem[{Ossewaarde and Gulenc(2020)}]{ossewaarde2020national}
Ossewaarde, M.; and Gulenc, E. 2020.
\newblock National varieties of artificial intelligence discourses: Myth, utopianism, and solutionism in West European policy expectations.
\newblock \emph{Computer}, 53(11): 53--61.

\bibitem[{Pepermans and Maeseele(2016)}]{politicization_climate_change}
Pepermans, Y.; and Maeseele, P. 2016.
\newblock The politicization of climate change: problem or solution?
\newblock \emph{WIREs Climate Change}, 7(4): 478--485.

\bibitem[{Peterson and Muñoz(2022)}]{Stick_To_Sports}
Peterson, E.; and Muñoz, M. 2022.
\newblock “Stick to Sports”: Evidence from Sports Media on the Origins and Consequences of Newly Politicized Attitudes.
\newblock \emph{Political Communication}, 39(4): 454--474.

\bibitem[{Rajadesingan, Budak, and Resnick(2021)}]{rajadesingan2021political}
Rajadesingan, A.; Budak, C.; and Resnick, P. 2021.
\newblock Political discussion is abundant in non-political subreddits (and less toxic).
\newblock In \emph{Proceedings of the Fifteenth International AAAI Conference on Web and Social Media}, volume~15.

\bibitem[{Schaffner and Roche(2016)}]{schaffner2016misinformation}
Schaffner, B.~F.; and Roche, C. 2016.
\newblock Misinformation and motivated reasoning: Responses to economic news in a politicized environment.
\newblock \emph{Public Opinion Quarterly}, 81(1): 86--110.

\bibitem[{Souza, Nogueira, and Lotufo(2020)}]{souza2020bertimbau}
Souza, F.; Nogueira, R.; and Lotufo, R. 2020.
\newblock BERTimbau: Pretrained BERT Models for Brazilian Portuguese.
\newblock In Cerri, R.; and Prati, R.~C., eds., \emph{Intelligent Systems}, 403--417. Cham: Springer International Publishing.
\newblock ISBN 978-3-030-61377-8.

\bibitem[{Sun and Loparo(2019)}]{sun2019topic}
Sun, Y.; and Loparo, K. 2019.
\newblock Topic shift detection in online discussions using structural context.
\newblock In \emph{2019 IEEE 43rd Annual Computer Software and Applications Conference (COMPSAC)}, volume~1, 948--949. IEEE.

\bibitem[{Taber, Cann, and Kucsova(2009)}]{taber2009motivated}
Taber, C.~S.; Cann, D.; and Kucsova, S. 2009.
\newblock The motivated processing of political arguments.
\newblock \emph{Political Behavior}, 31: 137--155.

\bibitem[{Tumasjan et~al.(2011)Tumasjan, Sprenger, Sandner, and Welpe}]{election_forecasts}
Tumasjan, A.; Sprenger, T.~O.; Sandner, P.~G.; and Welpe, I.~M. 2011.
\newblock Election Forecasts With Twitter: How 140 Characters Reflect the Political Landscape.
\newblock \emph{Soc. Sci. Comput. Rev.}, 29(4): 402–418.

\bibitem[{Wang, Zhang, and Liu(2017)}]{wang2017sentiment}
Wang, Y.; Zhang, Y.; and Liu, B. 2017.
\newblock Sentiment lexicon expansion based on neural pu learning, double dictionary lookup, and polarity association.
\newblock In \emph{Proceedings of the 2017 Conference on Empirical Methods in Natural Language Processing}.

\bibitem[{Weber, Garimella, and Borra(2013)}]{weber2013inferring}
Weber, I.; Garimella, V. R.~K.; and Borra, E. 2013.
\newblock Inferring audience partisanship for youtube videos.
\newblock In \emph{Proceedings of the 22nd Int'l Conference on World Wide Web}, 43--44.

\bibitem[{Wiesner(2021)}]{wiesner2021rethinking}
Wiesner, C. 2021.
\newblock \emph{Rethinking Politicisation in Politics, Sociology and International Relations}.
\newblock Palgrave Studies in European Political Sociology. Springer International Publishing.

\bibitem[{Wojcieszak and Mutz(2009)}]{wojcieszak2009online}
Wojcieszak, M.~E.; and Mutz, D.~C. 2009.
\newblock Online groups and political discourse: Do online discussion spaces facilitate exposure to political disagreement?
\newblock \emph{Journal of communication}, 59(1): 40--56.

\bibitem[{Wright(1998)}]{politicization_culture}
Wright, S. 1998.
\newblock The Politicization of 'Culture'.
\newblock \emph{Anthropology Today}, 14: 7.

\bibitem[{Zembylas, Loukaidis, and Antoniou(2019)}]{politicization_religion}
Zembylas, M.; Loukaidis, L.; and Antoniou, M. 2019.
\newblock The politicisation and securitisation of religious education in Greek–Cypriot schools.
\newblock \emph{European Educational Research Journal}, 18(1): 69--84.

\bibitem[{Zhao and Valentini(2022)}]{zhao2022navigating}
Zhao, H.; and Valentini, C. 2022.
\newblock Navigating turbulent political waters: From corporate political advocacy to scansis in the case of NBA-China crisis.
\newblock \emph{Journal of Public Relations Research}, 34(1-2): 64--87.

\bibitem[{Zürn(2019)}]{politicization_compared}
Zürn, M. 2019.
\newblock Politicization compared: at national, European, and global levels.
\newblock \emph{Journal of European Public Policy}, 26(7): 977--995.

\end{thebibliography}

\appendix

\section{Appendix A - Collected News Sources}
\label{appendix:news_sources}
\begin{table}[H]
\resizebox{\columnwidth}{!}{
\begin{tabular}{llc}
\toprule
\textbf{Channel}    & \textbf{Reach} & \textbf{Platform With Most Followers} \\
\midrule
g1                  & 14.8 mi        & Twitter           \\
VEJA                & 9.1 mi         & Twitter           \\
Folha de S.Paulo    & 8.8 mi         & Twitter           \\
Estadão             & 7.5 mi         & Twitter           \\
Jornal O Globo      & 7.3 mi         & Twitter           \\
Jovem Pan News      & 7.3 mi         & YouTube           \\
ge                  & 6.3 mi         & Twitter           \\
Globo News          & 5.6 mi         & Twitter           \\
UOL Notícias        & 5.2 mi         & Twitter           \\
ESPN Brasil         & 5.2 mi         & Twitter           \\
R7                  & 5.1 mi         & Twitter           \\
CNN Brasil          & 4.0 mi         & YouTube           \\
Jornal da Record    & 3.9 mi         & Youtube           \\
Metrópoles          & 3.6 mi         & TikTok            \\
Pânico Jovem Pan    & 3.6 mi         & YouTube           \\
BBC News Brasil     & 3.4 mi         & Twitter           \\
UOL                 & 3.4 mi         & YouTube           \\
Valor Econômico     & 2.6 mi         & Twitter           \\
Revista Oeste       & 1.3 mi         & YouTube           \\
GZH                 & 1.1 mi         & Twitter           \\
Correio Braziliense & 0.9 mi         & Twitter           \\
O TEMPO             & 0.5 mi         & YouTube           \\
Estado de Minas     & 0.5 mi         & Twitter           \\
A TARDE             & 0.5 mi         & Twitter           \\
SuperesportesMG     & 0.2 mi         & Twitter           \\  \bottomrule
\end{tabular}
}
\end{table}
\end{document}